\documentclass[10pt,twocolumn,superscriptaddress,amssymb,amsmath,aps,prd]{revtex4-2}
\usepackage{graphicx}

\usepackage{bbm}

\usepackage[normalem]{ulem}

\usepackage{color}

\usepackage{multirow}

\begin{document}


\title{Light deflection in unified gravity and measurable deviation from general relativity\\ in the second post-Newtonian order}
\date{September 16, 2025}
\author{Mikko Partanen}
\affiliation{Photonics Group, Department of Electronics and Nanoengineering, 
Aalto University, P.O. Box 13500, 00076 Aalto, Finland}
\author{Jukka Tulkki}
\affiliation{Engineered Nanosystems Group, School of Science, Aalto University, 
P.O. Box 12200, 00076 Aalto, Finland}

\begin{abstract}
Light does not travel in a perfectly straight line when it passes near massive objects. In this work, we calculate the gravitational deflection of light using the gauge theory of unified gravity [Rep.~Prog.~Phys.~\textbf{88}, 057802 (2025)], formulated as an extension of the Standard Model. The nonlinear graviton--graviton interaction is accounted for in the lowest order. The dynamical equations of light in external gravitational field in unified gravity are essentially different from the dynamical equations obtained using the curved metric of general relativity. The goal of the present work is to compare the predictions of unified gravity and general relativity for the gravitational deflection of light. Since the calculation of the gravitational deflection of light in unified gravity is fundamentally different from the pertinent calculation in general relativity, we devote ample space for approximations used. The deflection angles obtained from unified gravity and general relativity are equal in the first post-Newtonian (PN) order, and agree with previous experiments available for the Sun. However, the second PN order terms of unified gravity reveal measurable relative differences of $7/30\approx23.3$\% and $4/15\approx26.7$\% for the out-of-plane and in-plane polarizations, respectively, in comparison with the polarization-independent value of general relativity. Therefore, we expect that experimentally differentiating between the two theories will become possible in the near future. For future experiments, we also make calculations for other astrophysical objects, such as for a neutron star, which is in the limit of the PN approximation. The application of unified gravity to black holes, for which the PN approximation is not valid, is left as a topic of further work.
\end{abstract}

\maketitle


\section{Introduction}
\vspace{-0.3cm}

Gravitational lensing \cite{Saha2024,Einstein1936,Turyshev2009a} was first suggested within Newtonian physics by describing light as particles attracted by gravity \cite{Soldner1801}. Newton's framework predicts only half the amount of deflection that we observe and this result remains the same in special relativity. It was Einstein’s general theory of relativity (GR) that provided the correct explanation: mass does not just pull on objects—it warps the very fabric of spacetime \cite{Einstein1916}. Light follows the curvature of spacetime, bending more than Newtonian gravity alone would allow. This bending was famously confirmed during the 1919 solar eclipse \cite{Dyson1920}, offering another major experimental validation of Einstein’s theory, in addition to the previously discovered perihelion precession of Mercury \cite{Einstein1915,Verrier1859}.

\begin{figure}
\centering
\includegraphics[width=0.98\columnwidth]{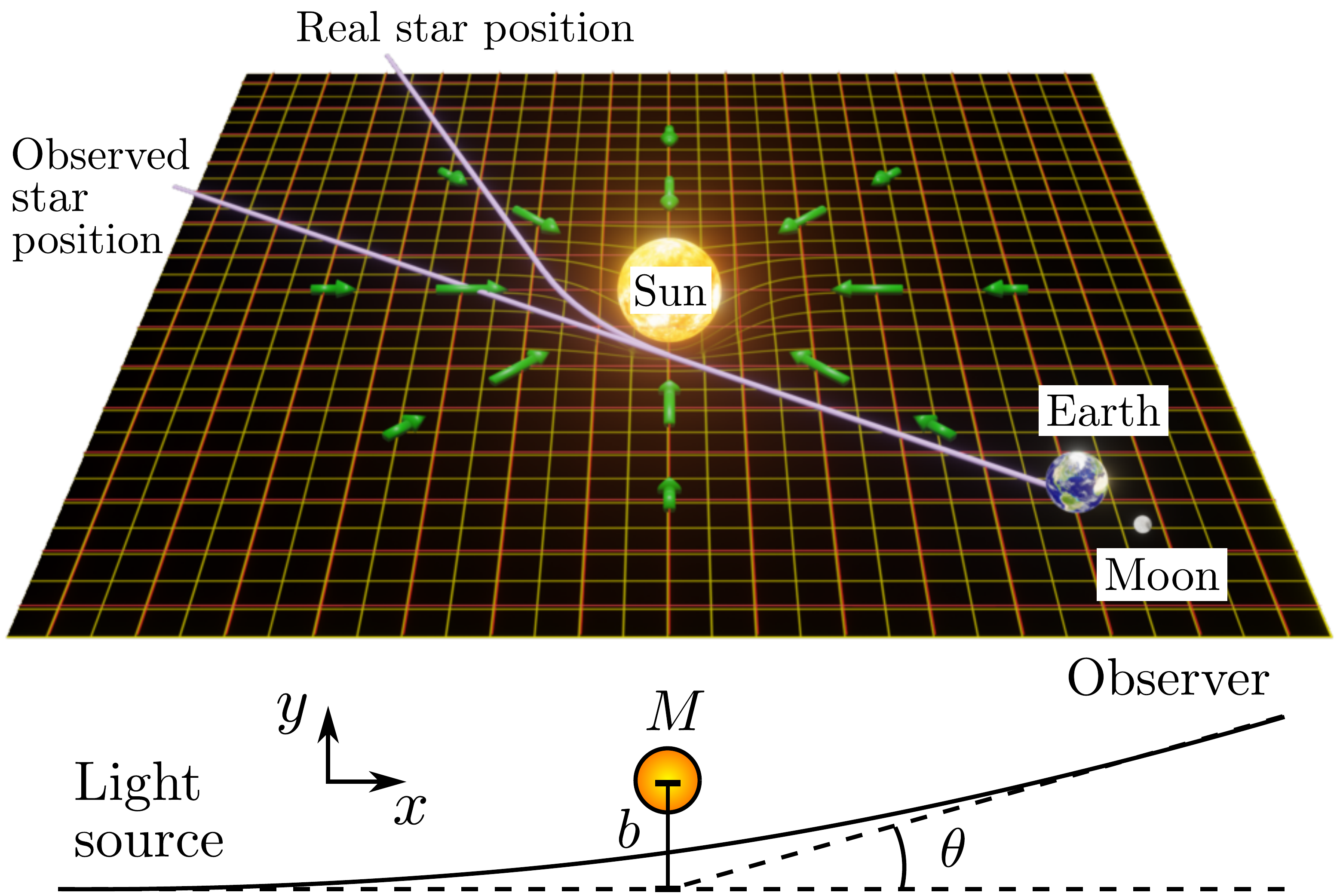}
\caption{\label{fig:illustration}
Illustration of the gravitational deflection of light by a massive astrophysical object, such as the Sun. Light is deflected by an angle $\theta$ due to the astrophysical object of mass $M$. The impact parameter $b$ is the closest distance of the mass and the initial light ray in the absence of deflection.}
\end{figure}

In this work, we study the gravitational deflection of light, illustrated in Fig.~\ref{fig:illustration}, using the gauge theory of unified gravity (UG) \cite{Partanen2025a}. Originally, the theory was presented in two formulations based on different geometric conditions. One of the geometric conditions leads to teleparallel equivalent of GR \cite{Partanen2025a,Bahamonde2023a,Aldrovandi2012,Maluf2013}. In this work, we use the Minkowski metric formulation, which preserves the four U(1) gauge symmetries of the theory and slightly deviates from GR already in the classical physics regime. Accordingly, the abbreviation UG in this work always stands for the Minkowski metric formulation of the theory. Unlike parametric modifications of GR \cite{Brans1961,Capozziello2007,Capozziello2010,deRham2011,Jimenez2018b,Soares2025}, UG contains only known physical constants. We show how UG can be used to calculate the gravitational field and to explain the observable deflection of light near massive objects. The field equations of UG are written in a global Minkowski frame, and the gravity gauge field appears as a conventional field together with all other fields in these equations. Therefore, in contrast to GR, the effect of gravity in UG is not hidden in the spacetime curvature described by the metric. Thus, the foundations of the theories are fundamentally different. Despite this difference, we show how perfect agreement between UG and GR is found for the gravitational deflection angle of light in the first post-Newtonian (1PN) order. This limit is valid in almost all cases of astrophysical interest apart from the physics of black holes. In general, the deflection of light obtained from UG differs from that predicted by GR. To quantify this difference, we calculate the second post-Newtonian (2PN) order deflection angle of light. This quantity is expected to be measurable in high-precision astrophysical experiments in the near future \cite{Turyshev2007,Turyshev2009b}. Other key gravitational effects, the perihelion precession of planetary orbits, and gravitational redshift are investigated using UG in separate works \cite{Partanen2025d,Partanen2025e}. These gravitational effects have been chosen because they offer an easy comparison of the predictive capabilities of UG and GR. Comparison with the state-of-the-art experiments on light bending in the strong gravitational fields of black holes has been left as a topic of further work.

\section{Unified gravity gauge field}
\vspace{-0.3cm}

We start our study with the solution of the gravity gauge field of UG for a point mass. In UG, we assume a global Minkowski frame with Cartesian coordinates $x^\mu=(ct,x,y,z)$, where $c$ is the speed of light in vacuum and in zero gravitational potential. Accordingly, we use the Minkowski metric tensor $\eta_{\mu\nu}$, given by the diagonal components $\eta_{00}=1$ and $\eta_{xx}=\eta_{yy}=\eta_{zz}=-1$.

The gravitational field equation of UG in the harmonic gauge, $P^{\mu\nu,\rho\sigma}\partial_\rho H_{\mu\nu}=0$, is given including the nonlinear graviton--graviton interaction studied in Ref.~\cite{Partanen2025f} by
\begin{equation}
 P^{\mu\nu,\rho\sigma}\partial^2H_{\rho\sigma}-P^{\sigma\lambda,\rho\mu\nu,\alpha\beta\gamma}\partial_\rho(H_{\sigma\lambda}\partial_\alpha H_{\beta\gamma})=-\kappa T^{\mu\nu}.
 \label{eq:UGgravitygeneral}
\end{equation}
Here $H_{\rho\sigma}$ is the gravity gauge field of UG, $\partial^2=\partial^\rho\partial_\rho$ is the d'Alembert operator, and $\kappa=8\pi G/c^4$ is Einstein's constant, in which $G$ is the gravitational constant. The source of gravity, $T^{\mu\nu}$, on the right in Eq.~\eqref{eq:UGgravitygeneral}, is the total stress-energy-momentum tensor of all fields and matter. The coefficients $P^{\mu\nu,\rho\sigma}$ and $P^{\sigma\lambda,\rho\mu\nu,\alpha\beta\gamma}$ are given in terms of the Minkowski metric tensor $\eta^{\mu\nu}$ as presented in Appendix~\ref{apx:coefficients}.

The stress-energy-momentum source term can be separated into the part, $T_\mathrm{m}^{\mu\nu}$, associated with matter and vector gauge fields, such as the electromagnetic field, and into the gravity part, $T_\mathrm{g}^{\mu\nu}$, as
\begin{equation}
 T^{\mu\nu}=T_\mathrm{m}^{\mu\nu}+T_\mathrm{g}^{\mu\nu},
\end{equation}
The gravity part of the stress-energy-momentum tensor is given in terms of the gravity gauge field $H_{\mu\nu}$ and the coefficients $P^{\sigma\lambda,\rho\mu\nu,\alpha\beta\gamma}$ by
\begin{align}
 T_\mathrm{g}^{\mu\nu}
 &=\frac{1}{2\kappa}P^{\mu\nu,\rho\sigma\lambda,\alpha\beta\gamma}\partial_\rho H_{\sigma\lambda}\partial_\alpha H_{\beta\gamma}.
 \label{eq:Tg}
\end{align}

We consider the solution of Eq.~\eqref{eq:UGgravitygeneral} for the stress-energy-momentum tensor of a point mass $M$ located at the origin, given by \cite{Landau1989}
\begin{equation}
 T_\mathrm{m}^{\mu\nu}=Mc^2\delta(\mathbf{r})\delta_0^\mu\delta_0^\nu.
 \label{eq:SEM}
\end{equation}
Here $\delta(\mathbf{r})$ is the Dirac delta function in the three-dimensional space coordinates $\mathbf{r}=(x,y,z)$, and $\delta_\nu^\mu$ is the Kronecker delta.

\subsection{Solution to the linearized field equation of gravity}
\vspace{-0.3cm}

For the solution of the nonlinear field equation of gravity in Eq.~\eqref{eq:UGgravitygeneral}, we use the iterative approach presented in Ref.~\cite{Partanen2025f}. First, we study the solution of the linearized equation, which is sufficient for the study of 1PN order effects. This equation is given by
\begin{equation}
 P^{\mu\nu,\rho\sigma}\partial^2H_{\rho\sigma}^{(1)}=-\kappa T_\mathrm{m}^{\mu\nu}.
 \label{eq:UGgravity0}
\end{equation}
The exact gravity gauge field solution to the linearized equation of gravity in Eq.~\eqref{eq:UGgravity0} for the stress-energy-momentum tensor in Eq.~\eqref{eq:SEM} is denoted by $H_{\mu\nu}^{(1)}$. It is straightforward to calculate and given by
\begin{equation}
 H_{\mu\nu}^{(1)}=\frac{\Phi}{c^2}\delta_{\mu\nu},
\hspace{1cm}\Phi=-\frac{GM}{r}.
 \label{eq:H0}
\end{equation}
Here $\Phi$ is the Newtonian gravitational potential satisfying Poisson's equation $\nabla^2\Phi=4\pi GM\delta(\mathbf{r})$, where $\nabla=(\partial_x,\partial_y,\partial_z)$ is the three-dimensional gradient operator, and $r=|\mathbf{r}|=\sqrt{x^2+y^2+z^2}$. The constants of integration have been set to zero by assuming that the gravitational field vanishes at infinity.

\subsection{2PN order solution to the nonlinear field equation of gravity}
\vspace{-0.3cm}

The 2PN order solution to the nonlinear field equation of gravity in Eq.~\eqref{eq:UGgravitygeneral} is obtained by approximating the nonlinear terms of this equation by their values calculated using the first-order solution $H_{\rho\sigma}^{(1)}$. Accordingly, Eq.~\eqref{eq:UGgravitygeneral} becomes
\begin{align}
 &P^{\mu\nu,\rho\sigma}\partial^2H_{\rho\sigma}^{(2)}-P^{\sigma\lambda,\rho\mu\nu,\alpha\beta\gamma}\partial_\rho(H_{\sigma\lambda}^{(1)}\partial_\alpha H_{\beta\gamma}^{(1)})\nonumber\\
 &=-\kappa(T_\mathrm{m}^{\mu\nu}+T_\mathrm{g}^{(1)\mu\nu}).
 \label{eq:UGgravity1}
\end{align}
Here $T_\mathrm{g}^{(1)\mu\nu}$ is the stress-energy-momentum tensor of gravity calculated using Eq.~\eqref{eq:Tg} for $H_{\rho\sigma}^{(1)}$. The solution to Eq.~\eqref{eq:UGgravity1}, denoted by $H_{\mu\nu}^{(2)}$, is given by
\begin{equation}
 H_{\mu\nu}^{(2)}= \Big(\frac{\Phi}{c^2}+\frac{\Phi^2}{c^4}\Big)\delta_{\mu\nu}
 -\frac{\Phi^2}{2c^4}\delta_{\mu0}\delta_{\nu0}
 -\frac{\Phi^2}{2c^4}\frac{r^ir^j}{r^2}\delta_{\mu i}\delta_{\nu j}.
 \label{eq:H1}
\end{equation}
Here the Latin indices range over the three spatial dimensions. The iterative approach can be continued to calculate higher-order contributions to $H_{\mu\nu}$. In the present work, we limit to the 2PN order solution in Eq.~\eqref{eq:H1}. This is a good approximation for weak fields with $|\Phi/c^2|\ll 1$ and its accuracy is sufficient to model previous weak-field experiments on gravity.

\section{Gravitational lensing from the dynamical equations}
\vspace{-0.3cm}

In UG, all dynamical equations are written in the global Minkowski frame, and the gravity gauge field appears explicitly in these equations. This is a fundamentally different starting point in comparison with GR, where gravitational coupling is only implicitly described by the metric \cite{Misner1973}. Next, we explicitly show that the deflection angles of light calculated from the dynamical equations of UG and GR agree, in the weak field limit, but differ when higher-order terms are considered.

The dynamical equation of the electromagnetic four-potential $A^\mu$ in UG in the absence of electric charges is given in the Feynman gauge, $\partial_\mu A^\mu=0$, as \cite{Partanen2025a,Partanen2025b}
\begin{align}
 \partial^2A^\sigma+P^{\mu\nu,\rho\sigma,\eta\lambda}\partial_\rho(H_{\mu\nu}\partial_\eta A_\lambda)=0.
 \label{eq:MaxwellUG}
\end{align}
Here the coefficients $P^{\mu\nu,\rho\sigma,\eta\lambda}$ in the gravity coupling term of UG are given in terms of the Minkowski metric tensor $\eta^{\mu\nu}$ as presented in Appendix~\ref{apx:coefficients}.

\subsection{Obtaining the refractive index profile}
\vspace{-0.3cm}

We assume that the propagation of light takes place in the $xy$ plane and light is initially propagating parallel to the $x$ axis. As conventional, we can safely neglect the small gravitational field produced by the light itself \cite{Saha2024}. For fixing the residual gauge degrees of freedom, we assume the electromagnetic four-potential $A^\mu=(A^0,\mathbf{A})$ of incident light in the radiation gauge \cite{Jackson1999} with $A^0=0$ and $\nabla\cdot\mathbf{A}=0$. For out-of-plane polarization of light parallel to the $z$ axis, the four-potential at $z=0$ is then of the form $A_\mathrm{s}^\mu(t,x,y)=[0,0,0,A^z(t,x,y)]$. Substituting the four-potential $A_\mathrm{s}^\mu$ and the second-order solution of the gravity gauge field, $H_{\mu\nu}^{(2)}$ from Eq.~\eqref{eq:H1}, into the dynamical equation of light in Eq.~\eqref{eq:MaxwellUG}, and accounting for the time independence of the Newtonian potential, we obtain after technical summation over repeated indices
\begin{align}
 &\frac{1}{c^2}\Big(1\!-\!\frac{2\Phi}{c^2}\!-\!\frac{\Phi^2}{c^4}\Big)\frac{\partial^2A^z}{\partial t^2}
 -\Big[1\!+\!\frac{2\Phi}{c^2}\!+\!\Big(1\!+\!\frac{y^2}{r^2}\Big)\frac{\Phi^2}{c^4}\Big]\frac{\partial^2A^z}{\partial x^2}\nonumber\\
 &\!-\!\Big[1\!+\!\frac{2\Phi}{c^2}\!+\!\Big(1\!+\!\frac{x^2}{r^2}\Big)\frac{\Phi^2}{c^4}\Big]\frac{\partial^2A^z}{\partial y^2}
 \!+\!\frac{2xy}{r^2}\frac{\Phi^2}{c^4}\frac{\partial^2A^z}{\partial x\partial y}\nonumber\\
 &\!-\!\frac{2}{c^2}\Big(1+\frac{3\Phi}{2c^2}\Big)\Big(\frac{\partial\Phi}{\partial x}\frac{\partial A^z}{\partial x}+\frac{\partial\Phi}{\partial y}\frac{\partial A^z}{\partial y}\Big)\!=\!0.
 \label{eq:Az}
\end{align}
Next, we consider the in-plane polarization of light, $A_\mathrm{p}^ \mu(t,x,y)=[A^0(t,x,y),A^x(t,x,y),A^y(t,x,y),0]$. Assuming that the deflection angle is small, the component $A^y(t,x,y)$ dominates over $A^0(t,x,y)$ and $A^x(t,x,y)$. Using Eq.~\eqref{eq:MaxwellUG} at $z=0$ for $\mu=y$, we obtain
\begin{align}
 &\frac{1}{c^2}\Big[1\!-\!\frac{2\Phi}{c^2}\!-\!\Big(1\!+\!\frac{y^2}{r^2}\Big)\frac{\Phi^2}{c^4}\Big]\frac{\partial^2A^y}{\partial t^2}
 -\Big(1\!+\!\frac{2\Phi}{c^2}\!+\!\frac{\Phi^2}{c^4}\Big)\nonumber\\
 &\times\Big(\frac{\partial^2A^y}{\partial x^2}
 \!+\frac{\partial^2 A^y}{\partial y^2}\Big)\!-\!\frac{xy}{r^2}\frac{\Phi^2}{c^6}\frac{\partial^2A^x}{\partial t^2}
 \!-\!\frac{2}{c^2}\Big(1+\frac{3\Phi}{2c^2}\Big)\frac{\partial\Phi}{\partial x}\nonumber\\
 &\times\Big(\frac{\partial A^y}{\partial x}-\frac{\partial A^x}{\partial y}\Big)\!=\!0.
 \label{eq:Ay}
\end{align}

Equations \eqref{eq:Az} and \eqref{eq:Ay} are componentwise wave equations of light in the gravitational lens as obtained in UG. The last terms of Eqs.~\eqref{eq:Az} and \eqref{eq:Ay} are proportional to the first derivatives of the vector potential components. Therefore, these terms describe attenuation and amplification of the field and do not contribute to the speed of light. The speed of light generally depends on the direction, and we obtain an effective permittivity tensor for the gravitational lens. The speed of light in the gravitational lens, $c'=c/n$, is determined by the factors of the first three terms of Eq.~\eqref{eq:Az} and the factors of the first two terms of Eq.~\eqref{eq:Ay}. The corresponding refractive indices for the two polarizations are given by
\begin{align}
 n_\mathrm{s}^x &=\sqrt{\frac{1\!-\!\frac{2\Phi}{c^2}\!-\!\frac{\Phi^2}{c^4}}{1\!+\!\frac{2\Phi}{c^2}\!+\!\big(1\!+\!\frac{y^2}{r^2}\big)\frac{\Phi^2}{c^4}}}
 \approx 1\!-\!\frac{2\Phi}{c^2}\!+\!\Big(1\!-\!\frac{y^2}{2r^2}\Big)\frac{\Phi^2}{c^4},\nonumber\\
 n_\mathrm{s}^y &=\sqrt{\frac{1\!-\!\frac{2\Phi}{c^2}\!-\!\frac{\Phi^2}{c^4}}{1\!+\!\frac{2\Phi}{c^2}\!+\!\big(1\!+\!\frac{x^{2\raise0.8ex\hbox{}}}{r^2}\big)\frac{\Phi^2}{c^4}}}
 \approx 1\!-\!\frac{2\Phi}{c^2}\!+\!\Big(1\!-\!\frac{x^2}{2r^2}\Big)\frac{\Phi^2}{c^4},\nonumber\\
 n_\mathrm{p}^x &\!=\!n_\mathrm{p}^y\!=\!\sqrt{\frac{1\!-\!\frac{2\Phi}{c^2}\!-\!\big(1\!+\!\frac{y^2}{r^2}\big)\frac{\Phi^2}{c^4}}{1\!+\!\frac{2\Phi}{c^2}\!+\!\frac{\Phi^2}{c^4}}}
 \!\approx 1\!-\!\frac{2\Phi}{c^2}\!+\!\Big(1\!-\!\frac{y^2}{2r^2}\Big)\frac{\Phi^2}{c^4}.
 \label{eq:nA}
\end{align}
The last forms of Eq.~\eqref{eq:nA} are obtained by taking the Taylor series in powers of $\Phi$ and truncating them after the second-order terms. The nonequality of $n_\mathrm{s}^x$ and $n_\mathrm{s}^y$ shows that the gravitational lens acts as an anisotropic medium for the out-of-plane polarization. Therefore, following the standard description of light in anisotropic media \cite{Landau1984}, we must use an effective refractive index for the out-of-plane polarization. The refractive indices for the two polarizations are then given by
\begin{equation}
 n_\mathrm{s}=\sqrt{\frac{x^2+y^2}{\big(\frac{x}{n_\mathrm{s}^{x\raise0.8ex\hbox{}}}\big)^2+\big(\frac{y}{n_\mathrm{s}^{y\raise0.8ex\hbox{}}}\big)^2}}
 \approx 1\!-\!\frac{2\Phi}{c^2}\!+\!\Big(1\!-\!\frac{y^2}{r^2}\!+\!\frac{y^4}{r^4}\Big)\frac{\Phi^2}{c^4},
 \label{eq:nAz}
\end{equation}
\begin{equation}
 n_\mathrm{p}=n_\mathrm{p}^x=n_\mathrm{p}^y
 \approx 1\!-\!\frac{2\Phi}{c^2}\!+\!\Big(1\!-\!\frac{y^2}{2r^2}\Big)\frac{\Phi^2}{c^4}.
 \label{eq:nAy}
\end{equation}
Equations \eqref{eq:nAz} and \eqref{eq:nAy} show that the two polarizations propagate at generally different velocities in the 2PN approximation to which our solution for $H_{\mu\nu}$ in Eq.~\eqref{eq:H1} is valid. Therefore, a gravitational lens can polarize light propagating through it. This effect is not predicted by GR as discussed further in Sec.~\ref{sec:comparison}. However, at very low frequencies, in which the ray optics approximation does not apply, GR also leads to polarization-dependent phenomena \cite{Oancea2020}.


\subsection{Ray-optics solution for the deflection angle}
\vspace{-0.3cm}

Next, we calculate the gravitational deflection angle of light following from the refractive index profiles determined above.
After obtaining the refractive index profiles in Eqs.~\eqref{eq:nAz} and \eqref{eq:nAy}, the problem of determining the gravitational deflection angle of light through the solution of Eqs.~\eqref{eq:Az} and \eqref{eq:Ay} can be reduced to a problem of classical ray optics. This approach is analogous to the corresponding study in GR \cite{Fischbach1980}. The ray optics approximation is known to be accurate when the wavelength is small in comparison with the length scales in the refractive index profile \cite{Jackson1999,Oancea2020}. For the slowly varying gravitational potential, this approximation is extremely accurate. Contributions of higher powers of $\Phi$ are obtained through an iterative process.

The definition of the velocity $c'$ as the difference of the space and time differentials $ds$ and $dt$ along the light curve, given by $c'=c/n=ds/dt$, leads to the total travel time to be given by an integral
\begin{align}
 t &=\int_{0}^{t} dt
 =\int_{s(0)}^{s(t)}\frac{ds}{c'}
 =\frac{1}{c}\int_{s(0)}^{s(t)}n(\mathbf{r})ds\nonumber\\
 &=\frac{1}{c}\int_{x(0)}^{x(t)}n[\mathbf{r}(x')]\Big|\frac{d\mathbf{r}(x')}{dx'}\Big|dx'
\end{align}
According to Fermat's principle, the physical path of light minimizes the travel time. Therefore, we obtain a variational problem, written as
\begin{equation}
 \delta\int_{x(0)}^{x(t)}L\Big(\mathbf{r},\frac{\partial\mathbf{r}}{\partial x'},x'\Big)dx'=0.
\end{equation}
Here the Lagrangian function $L$ is given by
\begin{equation}
 L\Big(\mathbf{r},\frac{d\mathbf{r}}{dx'},x'\Big)=n[\mathbf{r}(x')]\Big|\frac{d\mathbf{r}(x')}{dx'}\Big|.
\end{equation}
The Euler--Lagrange equation is written as
\begin{equation}
 \frac{d}{dx'}\Big[\frac{\partial L}{\partial(d\mathbf{r}/dx')}\Big]-\frac{\partial L}{\partial\mathbf{r}}=0.
 \label{eq:EulerLagrange}
\end{equation}
For the derivatives of the Lagrangian, we obtain
\begin{equation}
 \frac{\partial L}{\partial\mathbf{r}}=
 \Big|\frac{d\mathbf{r}}{dx'}\Big|\nabla n
 =\sqrt{1+\Big(\frac{dy}{dx'}\Big)^2}\nabla n,
 \label{eq:Lagrangianderivative1}
\end{equation}
\begin{equation}
 \frac{\partial L}{\partial(d\mathbf{r}/dx')}
 =n\frac{d\mathbf{r}/dx'}{|d\mathbf{r}/dx'|}
 =n\hat{\mathbf{s}}.
 \label{eq:Lagrangianderivative2}
\end{equation}
In Eq.~\eqref{eq:Lagrangianderivative1}, we have used the light curve parametrization $\mathbf{r}=[x',y(x'),0]$, where $y=y(x')$ defines the light curve. In Eq.~\eqref{eq:Lagrangianderivative2}, $\hat{\mathbf{s}}=(d\mathbf{r}/dx')/|d\mathbf{r}/dx'|$ is the unit tangent vector of the light curve. Substituting the Lagrangian derivatives from Eqs.~\eqref{eq:Lagrangianderivative1} and \eqref{eq:Lagrangianderivative2} into the Euler--Lagrange equation in Eq.~\eqref{eq:EulerLagrange}, after some algebra, we obtain
\begin{equation}
 \frac{d\hat{\mathbf{s}}}{dx'}
 =\sqrt{1+\Big(\frac{dy}{dx'}\Big)^2}\frac{\nabla n}{n}
 -\frac{dn}{ndx'}\frac{\Big(1,\dfrac{dy}{dx'},0\Big)}{\sqrt{1+\Big(\dfrac{dy}{dx'}\Big)^2}}.
 \label{eq:dsdx}
\end{equation}
Here $\nabla n=[\frac{\partial n(x',y)}{\partial x'},\frac{\partial n(x',y)}{\partial y},\frac{\partial n(x',y)}{\partial z}]|_{y=y(x')}$ is the gradient of the refractive index profile $n(x,y)$ evaluated at $x=x'$ and $y=y(x')$. The total derivative $dn/dx'$ is calculated from the refractive index profile $n(x,y)$ as $dn/dx'=\frac{\partial n(x',y)}{\partial x'}+\frac{\partial n(x',y)}{\partial y}\frac{dy}{dx'}$.

The unit tangent vector of the light curve has the polar coordinate representation $\hat{\mathbf{s}}=(\cos\theta,\sin\theta,0)$, where the angle $\theta$ is equal to the deflection angle. Therefore, the deflection angle can be determined from the $y$ component of $\hat{\mathbf{s}}$, equal to $\hat{\mathbf{y}}\cdot\hat{\mathbf{s}}=\sin\theta$, where $\hat{\mathbf{y}}$ is the unit vector parallel to the $y$ axis. For the deflection angle, we thus obtain
\begin{equation}
 \theta(x)=\arcsin[\hat{\mathbf{y}}\cdot\hat{\mathbf{s}}(x)]
 =\arcsin\Big(\int_{-\infty}^x\hat{\mathbf{y}}\cdot\frac{d\hat{\mathbf{s}}}{dx'}dx'\Big).
\end{equation}

The light curve $y=y(x)$ is unknown, but it can be obtained as an iterative solution to Eq.~\eqref{eq:dsdx}. The iteration is defined by \cite{Fischbach1980}
\begin{equation}
y^{(0)}(x)=-b,
\label{eq:iterative1}
\end{equation}
\begin{equation}
 \theta^{(n)}(x)=\arcsin\Big(\int_{-\infty}^x\hat{\mathbf{y}}\cdot\frac{d\hat{\mathbf{s}}}{dx'}\Big|_{y=y^{(n-1)}(x')}dx'\Big),
\end{equation}
\begin{equation}
 y^{(n)}(x)=y^{(0)}(x)+\int_{-\infty}^x\tan[\theta^{(n)}(x')]dx'.
 \label{eq:iterative3}
\end{equation}
Here $b$ is the impact parameter, which is the closest distance between the mass and the light ray in the absence of deflection as illustrated in Fig.~\ref{fig:illustration}. We use the unperturbed light ray, $y^{(0)}(x)=-b$, as the zeroth-order approximation.

Using the refractive index profiles of UG in Eqs.~\eqref{eq:nAz} and \eqref{eq:nAy} and the iterative approach, given in Eqs.~\eqref{eq:iterative1}--\eqref{eq:iterative3}, we obtain the UG prediction for the deflection angle for the out-of-plane and in-plane polarizations up to the 2PN order as
\begin{equation}
 \theta_\mathrm{UG,s}^{(2)}\approx\frac{4GM}{c^2b}+\frac{23\pi}{8}\Big(\frac{GM}{c^2b}\Big)^2,
 \label{eq:angleUGs}
\end{equation}
\begin{equation}
 \theta_\mathrm{UG,p}^{(2)}\approx\frac{4GM}{c^2b}+\frac{11\pi}{4}\Big(\frac{GM}{c^2b}\Big)^2.
 \label{eq:angleUGp}
\end{equation}

\begin{table*}
\setlength{\tabcolsep}{3.2pt}
\renewcommand{\arraystretch}{1.2}
\caption{\label{tbl:comparison}
Comparison of the 1PN and 2PN gravitational deflection angles of light for selected astrophysical objects as calculated using UG and GR. The selected astrophysical objects are Jupiter, the Sun, the massive star R136a1, and the neutron star RX J1856.5--3754. Here $\theta^{(1)}$ and $\theta^{(2)}$ are the 1PN and 2PN order predictions for the deflection angle, respectively. In the case of the neutron star, the 3PN and higher-order corrections are also important, but these corrections are not given in the table. The quantity $(\theta_\mathrm{UG}^{(2)}-\theta_\mathrm{GR}^{(2)})/\theta^{(1)}$ approximately describes the relative difference of UG and GR for the total deflection angle, but does not account for the higher-order terms important for the neutron star. The masses are given in units of the solar mass $M_\odot=1.988416\times10^{30}$ kg. The masses and radii are taken from Refs.~\cite{Prsa2016,Kalari2022,Brands2022,Potekhin2014}. An experimental value for the deflection angle is available for the Sun, agreeing with GR within the relative accuracy of $10^{-4}$ \cite{Titov2018}, and for Jupiter in the case of light passing farther from the surface of the planet \cite{Li2022,Fomalont2003}, agreeing with GR within the relative accuracy of about 4\%.}
\begin{tabular}{ccccccc}
   \hline\hline
   Astrophysical object & Mass ($M_\odot$) & Radius (m) & $\theta^{(1)}$ (deg) & $\theta_\mathrm{UG}^{(2)}\!-\!\theta^{(1)}$ (deg) & $\theta_\mathrm{GR}^{(2)}\!-\!\theta^{(1)}$ (deg) &  $(\theta_\mathrm{GR}^{(2)}\!-\!\theta_\mathrm{UG}^{(2)})/\theta^{(1)}$\\
   \hline
   Jupiter & $9.55\times10^{-4}$ & $7.149\times10^7$ & $4.520\times10^{-6}$ & s: $2.013\times10^{-13}$ & $2.625\times10^{-13}$ & s: $1.4\times10^{-8}$\\
   & & & & p: $1.925\times10^{-13}$ & & p: $1.5\times10^{-8}$\\
   Sun & 1 & $6.957\times10^8$ & $4.864\times10^{-4}$ & s: $2.331\times10^{-9}$ & $3.041\times10^{-9}$ & s: $1.5\times10^{-6}$\\
   & & & & p: $2.230\times10^{-9}$ & & p: $1.7\times10^{-6}$\\
   Massive star R136a1 & 196 & $2.97\times10^{10}$ & $2.23\times10^{-3}$ & s: $4.91\times10^{-8}$ & $6.41\times10^{-8}$ & s: $6.7\times10^{-6}$\\
   & & & & p: $4.70\times10^{-8}$ & & p: $7.7\times10^{-6}$\\
   Neutron star RX J1856.5--3754 & 1.5 & $12.1\times10^3$ & 42.1 & s: 17.3 & 22.6 & s: 0.126\\
   & & & & p: 16.6 & & p: 0.144\\
   \hline\hline
 \end{tabular}
\vspace{-0.2cm}
\end{table*}

For comparison, using the refractive index profile of the gravitational lens in GR, given in Sec.~\ref{sec:comparison} below, we obtain the well-known deflection angle of GR up to the 2PN order, given by \cite{Fischbach1980}
\begin{equation}
 \theta_\mathrm{GR}^{(2)}\approx\frac{4GM}{c^2b}+\frac{15\pi}{4}\Big(\frac{GM}{c^2b}\Big)^2.
 \label{eq:angleGR}
\end{equation}
Comparison of Eqs.~\eqref{eq:angleUGs}--\eqref{eq:angleGR} shows that the 1PN order terms of the deflection angle are identical between UG and GR. The 1PN order term accurately explains the previous measurements of the gravitational deflection of light by astrophysical objects, such as the Sun \cite{Dyson1920,Misner1973}. However, comparison of the 2PN order terms of Eqs.~\eqref{eq:angleUGs}--\eqref{eq:angleGR} shows notable differences in the prefactors. The 2PN order terms of UG for the out-of-plane and in-plane polarizations are $7/30\approx23.3$\% and $4/15\approx26.7$\% smaller than 2PN order term of GR, respectively. These differences mean that UG leads to a slightly smaller deflection of light than GR, in addition to the polarization dependence discussed above. This is the most significant result of the present work, and it is expected to be measurable in the near future. Especially, the proposed Laser Astrometric Test of Relativity (LATOR) experiment is designed specifically for this \cite{Turyshev2007,Turyshev2009b}.

Numerical comparison of UG and GR for the gravitational deflection of light in selected astrophysical objects is presented in Table \ref{tbl:comparison}. This comparison shows that the relative difference of the theories for the total deflection angle of light in planets is very small being of the order of $10^{-8}$ for Jupiter. In conventional stars, the relative difference is still small being of the order of $10^{-6}$, but increases significantly in the case of the higher gravitational field strength of a neutron star. In the case of neutron stars, higher-order terms of the deflection angle also become important, but the study of their effect is left as a topic of further work. In detailed analyses of the 2PN effects, environmental factors, such as line-of-sight large-scale structures, must be accounted for. This is beyond the scope of the present work.


\subsection{\label{sec:comparison}Comparison of the dynamical equations of light in UG and GR in the 2PN order}
\vspace{-0.3cm}

Next, we briefly compare the dynamical equations of light in UG and GR in the 2PN order. The dynamical equation of the electromagnetic field in GR is given by $\nabla_\nu F^{\mu\nu}=0$ \cite{Misner1973}, where $F^{\mu\nu}=\nabla^\mu A^\nu-\nabla^\nu A^\mu$ and $\nabla_\nu$ denotes the Levi-Civita coordinate-covariant derivative. For comparison with UG, we use the Lorenz gauge, written using the partial derivative instead of the coordinate-covariant derivative of GR as $\partial_\nu A^\nu=0$. The dynamical equation of light in GR for the out-of-plane polarization $A_\mathrm{s}^\mu(t,x,y)=[0,0,0,A^z(t,x,y)]$ in the Schwarzschild metric in isotropic Cartesian coordinates expanded up to the 2PN order in the $xy$ plane becomes
\begin{align}
 &\frac{1}{c^2}\Big(1-\frac{2\Phi}{c^2}+\frac{2\Phi^2}{c^4}\Big)\frac{\partial^2 A^z}{\partial t^2}-\Big(1+\frac{2\Phi}{c^2}+\frac{5\Phi^2}{2c^4}\Big)\nonumber\\
 &\times\Big(\frac{\partial^2A^z}{\partial x^2}+\frac{\partial^2A^z}{\partial y^2}\Big)
 +\frac{1}{4c^2}\Big(\frac{\partial\Phi^2}{\partial y}\frac{\partial A^z}{\partial y}
 +\frac{\partial\Phi^2}{\partial x}\frac{\partial A^z}{\partial x}\Big)=0.
 \label{eq:MaxwellGR1}
\end{align}
Correspondingly, for the in-plane polarization of light, $A_\mathrm{p}^ \mu(t,x,y)=[A^0(t,x,y),A^x(t,x,y),A^y(t,x,y),0]$, using $\nabla_\nu F^{\mu\nu}=0$ at $z=0$ for $\mu=y$, we obtain
\begin{align}
 &\frac{1}{c^2}\Big(1-\frac{2\Phi}{c^2}+\frac{2\Phi^2}{c^4}\Big)\frac{\partial^2 A^y}{\partial t^2}-\Big(1+\frac{2\Phi}{c^2}+\frac{5\Phi^2}{2c^4}\Big)\nonumber\\
 &\times\Big(\frac{\partial^2A^y}{\partial x^2}
 +\frac{\partial^2A^y}{\partial y^2}\Big)
 +\frac{1}{4c^4}\frac{\partial\Phi^2}{\partial x}\Big(\frac{\partial A^y}{\partial x}-\frac{\partial A^x}{\partial y}\Big)=0.
 \label{eq:MaxwellGR2}
\end{align}
The refractive index profiles of the gravitational lens in GR are then obtained from the coefficients of the second-order derivative terms of Eqs.~\eqref{eq:MaxwellGR1} and \eqref{eq:MaxwellGR2}. The refractive index profiles of GR are equal for the out-of-plane and in-plane polarizations, given by
\begin{equation}
 n_\mathrm{s}=n_\mathrm{p}=\sqrt{\frac{1\!-\!\frac{2\Phi}{c^2}\!+\!\frac{2\Phi^2}{c^4}}{1\!+\!\frac{2\Phi}{c^2}\!+\!\frac{5\Phi^2}{2c^4}}}
 \approx 1\!-\!\frac{2\Phi}{c^2}\!+\!\frac{7\Phi^2}{4c^4}.
 \label{eq:nGR}
\end{equation}
Comparison of the refractive index profiles of the gravitational lens in UG and GR in Eqs.~\eqref{eq:nAz}, \eqref{eq:nAy}, and \eqref{eq:nGR} shows that the refractive index profiles are equal in the 1PN order, but the 2PN order terms are \emph{substantially different}. The refractive index profile of GR depends on the Newtonian potential with constant prefactors, but the refractive index profile of UG has explicit dependencies on $y$ and $r$. Therefore, the refractive index profile of UG in Eqs.~\eqref{eq:nAz} and \eqref{eq:nAy} cannot be described by the most conventional constant parameters of the parametrized post-Newtonian (PPN) formalism \cite{Fischbach1980,Will2014}. This is a signature of the $4\times \mathrm{U(1)}$ gauge symmetry of UG and clarifies that UG is not a weak-field approximation of GR.

\section{Conclusion}
\vspace{-0.3cm}

We have demonstrated that UG can describe the gravitational deflection of light without a curved metric, used in GR. We have solved the gravity gauge field from the nonlinear equation of gravity in UG for a classical point mass and used it in the dynamical equation of light to calculate the effective refractive index profile and the gravitational deflection angle of light near massive objects.

The field equation of light in UG contains explicit coupling to the gravity gauge field in contrast to describing gravity through the metric as in GR. Furthermore, the refractive index profile of the gravitational lens in UG cannot be described by the most conventional constant PPN parameters. This clarifies that UG is not a weak-field approximation of GR. For the gravitational deflection of light, UG and GR agree with each other in the 1PN order. However, our calculation of the 2PN order contribution to the deflection angle of light using UG shows significant, polarization-dependent differences of $7/30\approx23.3$\% and $4/15\approx26.7$\% for the out-of-plane and in-plane polarizations, respectively, in comparison with the pertinent polarization-independent value of GR. We have also presented a numerical comparison of UG and GR for the 1PN and 2PN deflection angles of light in selected astrophysical objects.

Detailed measurements of gravitational lensing in the 2PN order \cite{Turyshev2007,Turyshev2009b,Ulbricht2020} are expected to enable experimentally differentiating between the predictions of UG and GR in the near future. The investigation of strong-field effects of black holes using UG requires direct methods for the solution of the nonlinear field equations of UG since the PN expansion no longer converges. It is left as a topic for further work.

\section*{Data availability statement}
\vspace{-0.3cm}
All data that support the findings of this study are included
within the article (and any supplementary files).

\begin{acknowledgments}
\vspace{-0.3cm}
This work has been funded by the Research Council of Finland under Contract No.~349971. We thank Holger Merlitz for useful discussions.
\end{acknowledgments}

\vspace{0.7cm}

\appendix

\section{\label{apx:coefficients}Coefficients}
\vspace{-0.3cm}

In this appendix, we present the definitions of several prefactors in Eqs.~\eqref{eq:UGgravitygeneral}, \eqref{eq:Tg}, \eqref{eq:UGgravity0}, \eqref{eq:UGgravity1}, and \eqref{eq:MaxwellUG}, given in terms of the Minkowski metric tensor $\eta^{\mu\nu}$. These definitions are written as
\begin{equation}
 P^{\mu\nu,\rho\sigma}=\frac{1}{2}(\eta^{\mu\sigma}\eta^{\rho\nu}+\eta^{\mu\rho}\eta^{\nu\sigma}-\eta^{\mu\nu}\eta^{\rho\sigma}),
\end{equation}
\begin{align}
 P^{\mu\nu,\sigma\rho\lambda,\beta\alpha\gamma}
 &=D^{\lambda\mu,\rho\sigma\nu,\alpha\beta\gamma}-D^{\lambda\mu,\rho\sigma\nu,\alpha\gamma\beta}\nonumber\\
 &\hspace{0.4cm}-D^{\sigma\mu,\rho\lambda\nu,\alpha\beta\gamma}+D^{\sigma\mu,\rho\lambda\nu,\alpha\gamma\beta},
\end{align}
\begin{align}
 D^{\lambda\mu,\rho\sigma\nu,\alpha\beta\gamma} &=\eta^{\lambda\mu}C^{\rho\sigma\nu,\alpha\beta\gamma}+\eta^{\lambda\nu}C^{\rho\sigma\mu,\alpha\beta\gamma}\nonumber\\
 &\hspace{0.4cm}-\frac{1}{2}\eta^{\mu\nu}C^{\rho\sigma\lambda,\alpha\beta\gamma},
\end{align}
\begin{align}
 C^{\rho\mu\nu,\alpha\beta\gamma} &=\frac{1}{2}(\eta^{\nu\alpha}\eta^{\mu\beta}\eta^{\rho\gamma}+\eta^{\mu\alpha}\eta^{\rho\beta}\eta^{\nu\gamma}-\eta^{\rho\alpha}\eta^{\nu\beta}\eta^{\mu\gamma})\nonumber\\
 &\hspace{0.4cm}+\eta^{\rho\mu}\eta^{\alpha\gamma}\eta^{\nu\beta}-\eta^{\rho\nu}\eta^{\alpha\gamma}\eta^{\mu\beta},
\end{align}
\begin{align}
 &P^{\mu\nu,\rho\sigma,\eta\lambda}
 =\eta^{\eta\sigma}\eta^{\lambda\mu}\eta^{\nu\rho}-\eta^{\eta\mu}\eta^{\lambda\sigma}\eta^{\nu\rho}-\eta^{\eta\rho}\eta^{\lambda\mu}\eta^{\nu\sigma}\nonumber\\
 &\hspace{0.3cm}+\eta^{\eta\mu}\eta^{\lambda\rho}\eta^{\nu\sigma}-\eta^{\mu\sigma}\eta^{\nu\lambda}\eta^{\rho\eta}+\eta^{\mu\sigma}\eta^{\nu\eta}\eta^{\rho\lambda}+\eta^{\mu\rho}\eta^{\nu\lambda}\eta^{\sigma\eta}\nonumber\\
 &\hspace{0.3cm}-\eta^{\mu\rho}\eta^{\nu\eta}\eta^{\sigma\lambda}-\eta^{\mu\nu}\eta^{\eta\sigma}\eta^{\lambda\rho}+\eta^{\mu\nu}\eta^{\eta\rho}\eta^{\lambda\sigma}.
\end{align}

\end{document}